\documentclass[aps,prb,superscriptaddress,twocolumn,reprint]{revtex4-1}
\usepackage{graphicx}
\usepackage{mathrsfs}
\usepackage{bm}
\usepackage{amsmath}
\usepackage{dcolumn}
\usepackage{epstopdf}
\usepackage{dsfont}
\usepackage{amssymb}
\usepackage{tabularx}
\usepackage{array}
\usepackage{color}
\usepackage[colorlinks=true, letterpaper=true, pdfstartview=FitV, linkcolor=blue, citecolor=blue, urlcolor=blue]{hyperref}

\begin{document}

\preprint{AIP/123-QED}

\title{Time reversal preserved spin valve and spin transistor based on unconventional $p$-wave  magnets}
\author{Ze-Yong Yuan}
\affiliation{School of Physics and Materials Science, Guangzhou University, Guangzhou 510006, China}
\author{Jun-Feng Liu}
\email{phjfliu@gzhu.edu.cn}
\affiliation{School of Physics and Materials Science, Guangzhou University,
	Guangzhou 510006, China}
\author{Pei-Hao Fu}
\email{phy.phfu@gmail.com}
\affiliation{School of Physics and Materials Science, Guangzhou University, Guangzhou 510006, China}
\author{Jun Wang}
\email{jwang@seu.edu.cn}
	\affiliation{Department of Physics, Southeast University, Nanjing 210096,
		China}	

\begin{abstract}
The anisotropic spin splitting in unconventional magnets opens new opportunities for realizing spintronic functionalities without relying on net magnetization or relativistic spin-orbit coupling.
Here, we propose a spin valve and a spin transistor based on unconventional $p$-wave magnets (UPMs).
The spin valve is realized in a junction where a normal metal is sandwiched between two UPMs whose exchange-field strength vectors are oriented transverse to the junction direction.
The conductance of such a device is governed by the spin alignment between two UPMs: when their strength vectors are parallel, the spin-state alignment enables efficient electron transmission, leading to a high-conductance state; in contrast, the antiparallel configuration suppresses the conductance owing to the opposite spin orientations.
Furthermore, the spin-valve can be extended to a spin transistor by replacing the central normal metal with another UPM with a longitudinally oriented strength vector and a perpendicular spin polarization axis.
The central UPM enables uniform spin precession with the same precession frequency for all transverse modes. Both devices can be electrically controlled by modulating the strength vectors of UPMs. These findings establish UPMs as a promising platform for developing spintronic devices without net magnetization or relativistic spin-orbit coupling.
\end{abstract}

\maketitle

\section{Introduction}\label{Introduction}

Unconventional magnets (UMs) \cite{ref1,ref2,ref3,ref4,ref5,ref6,ref7,ref8,ref9,ref10,zeng2026classification} represent the third distinct magnetic phase that breaks the traditional dichotomy between ferromagnetism \cite{Nobel1971} and antiferromagnetism \cite{RevModPhys.25.58}.
UMs are characterized by a directional spin-split Fermi surface that lifts the spin degeneracy analogously to ferromagnets while maintaining vanishing net magnetization in antiferromagnets.
Based on the symmetry of the Fermi surface in momentum space, unconventional magnets (UMs) are classified into two categories\cite{ref9}: even-parity magnetic orders (e.g., $d$-, $g$-, and $i$-wave types)\cite{ref7,ref10} and odd-parity magnetic orders (e.g., $p$-, $f$- and $h$-wave types)\cite{ref6,ref11,ref12,zk69-k6b2}.
Even-parity magnetic order, corresponding to the degree \(\ell = 0, 2, 4,\) or \(6\) of the exchange magnetic field polynomials with momentum space as variables, breaks time-reversal symmetry while preserving inversion symmetry. This results in spin-split energy bands that satisfy $E_\sigma(\mathbf{k})=E_\sigma(-\mathbf{k})$ but $E_\sigma(\mathbf{k}) \neq E_{-\sigma}(\mathbf{k})$. In contrast, odd-parity magnetic orders, corresponding to the degree \(\ell = 1, 3,\) or \(5\) of the exchangemagnetic field in momentum, preserve time-reversal symmetry but break inversion symmetry. Their band structure is characterized by the relations $E_\sigma(\mathbf{k}) \neq E_\sigma(-\mathbf{k})$ and $E_\sigma(\mathbf{k}) = E_{-\sigma}(-\mathbf{k})$ \cite{ref9,ref42}.

In altermagnets, the even-parity spin splitting underlies a variety of spintronic and transport phenomena \cite{ref4,ref5}, including giant magnetoresistance \cite{ref14}, spin-transfer torques \cite{ref40,ref41, han2024harnessingmagneticoctupolehall}, spin filtering \cite{samanta2024spinfilteringinsulatingaltermagnets, ref43,Lv2025Gate}, spin precession and spin transistor\cite{Liu2026Altermagnetic}, spin pumping effects \cite{PhysRevB.108.L140408}, non-linear transports \cite{Zhu2025}, light-matter interactions \cite{Werner2024High,Fajollahpour2025Light,fu2025floquetengineeringspintriplet,Fu2025Light}, magneto-optical effects \cite{ref39}, light-induced spin density \cite{fu2025floquetengineeringspintriplet, Fu2025Light}, non-Hermitian electronic responses \cite{PhysRevB.110.235401, Dash_2025}, strongly correlation in Mott insulators \cite{PhysRevMaterials.8.064403} and other novel phenomena \cite{zhang2026sliding,lei2025shear,guo2026hidden,PhysRevLett.134.136301,Chen_2025, Fukaya2025Crossed, Lu2025Engineering, Fukaya2025Josephson, Maeda2025Classification, Zhao2025Orientation, Chen2025Unconventional, Chen2024Enumeration, Liu2022Spin, Liu2025Different, Duan2025Antiferroelectric, Duan2025Antiferroelectric, Gu2025Ferroelectric, Zhu2024Observation,Li2025TwoDimensional, Peng2025Ferroelastic, Fang2025Edgetronics, Yang2025Altermagnetic}.
 With all the above-mentioned developments in both theory and experiments \cite{ref16,ref18,ref19,ref20,ref21,ref22,ref23,ref24,ref25}, altermagnets have been established as a versatile platform for generating and manipulating spin currents \cite{PhysRevB.108.L140408,ref26,ref27,ref28,ref29,ref30,ref31,ref33}, which can be realized in a wide range of representative candidate materials include RuF$_4$~\cite{ref35}, MnTe~\cite{ref36,ref37}, CoNb$_3$S$_6$~\cite{ref34}, and RuO$_2$~\cite{ref38}, although its altermagnetic nature is still under debate\cite{npjSpintronics.2.50,arXiv.2510.13781,PhysRevLett.133.176401,2025arXiv251101647T}.

Research on odd-parity unconventional magnets remains comparatively less explored, compared to their well-developed even-parity counterparts discussed above.
Particularly, UMs with $p$-wave symmetry, referred to as unconventional $p$-wave magnets (UPMs) \cite{ref10}, exhibit momentum-dependent spin splitting and nonrelativistic spin--momentum locking.
Thus, by effectively mimicking spin-orbit coupling \cite{Galitski2013Spin,Manchon2015New} and preserving time-reversal symmetry, UPMs provide an appealing platform for scalable spintronic applications~\cite{ref4,ref48,ref49,ref50}.
Extensive efforts have been devoted to exploring their normal-state properties \cite{ref60,ref61,ref62,ref63}, including tunneling magnetoresistance \cite{ref12,ref51}, orientation-dependent anomalous Hall effects \cite{ref52}, spin-current generation \cite{ref42}, and non-Hermitian responses \cite{ref64,ref65}.
Their superconducting counterparts have also attracted considerable attention, revealing rich phenomena such as tunneling spin Hall effects \cite{ref53}, transverse spin supercurrents \cite{ref55}, and orientation-dependent transport in UPM-superconductor hybrids \cite{Fukaya2025Josephson,ref54,ref56,ref57,ref59}, among other phenomena \cite{ref57,ref62, PhysRevB.111.054501}.
On the materials side, UPMs have been theoretically proposed in compounds such as Mn$_3$GaN and CeNiAsO~\cite{ref6}, and very recent experiments have reported their realization in thin flakes of  NiI$_2$~\cite{Song2025Electrical} and metallic Gd$_3$Ru$_4$Al$_{12}$~\cite{Yamada2025Metallic}.
Notably, the demonstrated ability to electrically switch the N\'eel vector in UPMs~\cite{ref66,ref67} further underscores their potential for spintronic memory applications~\cite{Song2025Electrical}.

Despite the above-mentioned advances, proposals for spintronic devices directly leveraging $p$-wave magnets remain largely unexplored.
Two prototypical spintronic architectures, the spin valve \cite{DalDin2024Antiferromagnetic,ref70} and the spin field-effect transistor (SFET) \cite{Ciorga_2025,WOS:000223375100042,Jiang2019}, offer ideal testbeds for realizing nonrelativistic, time-reversal-symmetric spin control.
In a conventional spin valve \cite{DalDin2024Antiferromagnetic,ref70}, two ferromagnetic layers can be tuned between parallel (\text{P}) and antiparallel (\text{AP}) magnetization states: conduction occurs only for spins aligned with the local magnetization, leading to high conductance in the \text{P} configuration and suppression in the \text{AP} configuration.
Switching between P and AP states typically requires an external magnetic field \cite{ref43}.
In contrast, an SFET exploits spin precession \cite{Ciorga_2025,WOS:000223375100042,Jiang2019}: as electrons travel through the channel, their spins precess due to spin-orbit coupling, and a gate voltage modulates the precession frequency, thereby controlling the conductance.
Such devices leverage the spin degree of freedom for information processing and storage, promising low power consumption, high speed, and dense integration \cite{ DalDin2024Antiferromagnetic,Ciorga_2025}.

In this work, we propose a \textit{time-reversal-symmetric} spin valve and a spin transistor based on UPMs.
As illustrated in Fig.~\ref{fig:1}, the junction functions as a spin valve when the two UPM electrodes possess transverse exchange-field strength vectors and the central region acts as a normal metal (zero exchange-field strength).
Under AP alignment, the conductance is strongly suppressed at low Fermi energies, whereas P alignment restores finite conductance.
This arises from the anisotropic spin splitting of the Fermi surfaces along the transverse direction: when the Fermi contours on both sides are fully separated in the momentum space, the junction behaves as an ideal spin valve.
Moreover, replacing the central normal region with another UPM whose strength vector points longitudinally perpendicular to those in the electrodes converts the device into a spin transistor.
The longitudinally split Fermi surface induces coherent spin precession across all transverse momenta, leading to oscillatory conductance analogous to the Datta-Das mechanism \cite{Datta1990Electronic}.
Since the magnitude and orientation of the UPM strength vectors can be effectively controlled via electrically tuning of the spin polarization \cite{Song2025Electrical}, the proposed spin valves and spin transistors provide a fully nonrelativistic route toward spintronic functionality without net magnetization or spin-orbit coupling. Compared with other unconventional magnets, the global shift of Fermi surfaces in a p-wave magnet allows for perfect spin precession in spin transistors. Meanwhile, the well-separated Fermi surfaces of p-wave magnets, which are also present in some unconventional antiferromagnets\cite{ref14}, enable high-contrast switching in spin valves.

\section{Model and Methods}\label{Theory and formulations}

\begin{figure}[tbp]
\centering
\includegraphics[width=\linewidth]{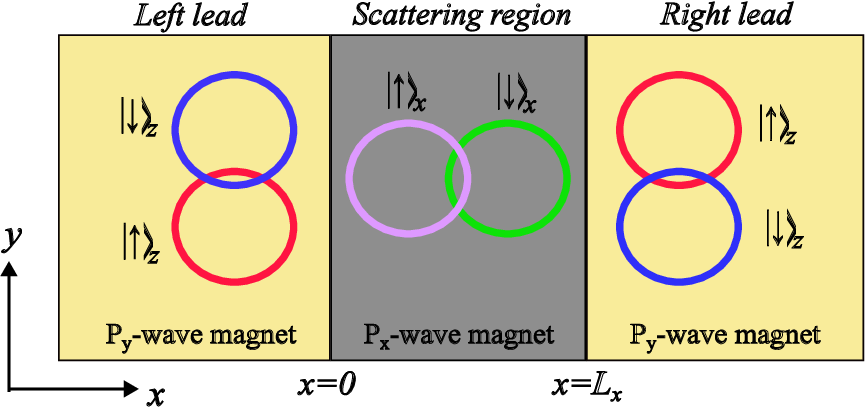}
\caption{Schematic of a UPM/UPM/UPM junction. In the left and right leads, the strength vectors of two UPMs are along the $y$ direction, inducing a splitting of the Fermi surface along the $k_y$ direction. And the spin polarization is along the $z$ direction. In the central UPM, the strength vector is along the $x$ direction and the spin polarization is along the $x$ direction, enabling a spin precession.}
\label{fig:1}
\end{figure}

As a unified model for a spin valve and a spin transistor, we consider a UPM/UPM/UPM trilayer junction, as shown in Fig. \ref{fig:1}.
For the left and right UPM leads, the strength vectors are along the $y$ direction and the spin polarization is along the $z$ direction. For the central UPM, the strength vector is along the $x$ direction and the spin polarization is along the $x$ direction. The tight-binding Hamiltonian for the junction is given by \cite{ref42}

\begin{equation}
\begin{aligned}
H&= - t_{0} \sum_{\bm{j}} \left(C_{\bm{j}}^{\dagger}C_{\bm{j}+\hat{x}}+C_{\bm{j}}^{\dagger}C_{\bm{j}+\hat{y}}+\text{h.c.}\right)+4t_0 \\
  &\quad + \sum_{\bm{j}} \left\{ it_{x}[\Theta(x)-\Theta(x-L_x)]C_{\bm{j}}^{\dagger}\sigma_{x}C_{\bm{j}+\hat{x}}+\text{h.c.}\right\} \\
  &\quad + \sum_{\bm{j}} \left\{i[t_{yL}\Theta(-x)+t_{yR}\Theta(x-L_x)]C_{\bm{j}}^{\dagger}\sigma_{z}C_{\bm{j}+\hat{y}}+\text{h.c.}\right\} \\
  &\quad +U_{0}\sum_{\bm{j}} [\Theta(x)-\Theta(x-L_x)]C_{\bm{j}}^{\dagger}C_{\bm{j}}+ \gamma \sum_{j_{x} \in \{0, L_{x}\},j_y}  C_{\bm{j}}^{\dagger}C_{\bm{j}},
\end{aligned}
\label{eq:1}
\end{equation}
where $C_{\bm{j}}^{\dagger}=(C_{\bm{j}\uparrow}^{\dagger}, C_{\bm{j}\downarrow}^{\dagger})$ are electron creation operators, $\bm{j} = (j_x, j_y)$ denotes the lattice coordinates, $\hat{x}$ ($\hat{y}$) is the unit vector in the $x$ ($y$) direction, and $\Theta(x)$ is the step function. Here $t_0$ denotes the nearest-neighbor hopping energy, $t_{yL}$ ($t_{yR}$) is the magnitude of the strength vector along the $y$ direction in the left (right) UPM lead, and $t_{x}$ is the magnitude of the strength vector along the $x$ direction in the central UPM. $\sigma_{x}$ and $\sigma_{z}$ are Pauli matrices in spin space. $U_0$ is the on-site energy in the central UPM region, and $\gamma$ represents the strength of the interfacial barriers.

For simplicity, the translational symmetry along the $y$ direction is assumed to be preserved and the transverse momentum $k_y$ is conserved. Within the Landauer-B\"{u}ttiker formalism, the conductance at a fixed Fermi energy $E_F$ is expressed in terms of the total transmission probability $T(k_y, E_F)$ \cite{Datta_1995},
\begin{equation}
G= G_0 \int  T(E_F,k_y) dk_y,
\label{eq:2}
\end{equation}
where $G_0=2e^2 L_y/(2\pi h)$, $L_y$ is the width of the system. The transmission probability can be calculated by the lattice Green function technique \cite{ref43,PhysRevB.73.075303,JAUHO_2003,Haug2008}
\begin{equation}
T(k_y, E_F)=\text{Tr}\left(\Gamma_LG^{r}\Gamma_RG^{a}\right).
\label{eq:3}
\end{equation}
Here,
\begin{equation}
    \Gamma_{L/R} = i\left[\Sigma_{L/R}^r - (\Sigma_{L/R}^{r})^{\dagger}\right]
    \label{eq:4}
\end{equation}
are the linewidth functions. The retarded Green function is
\begin{equation}
    G^{r}(E)=\left[E-H_{C}-\Sigma_{L}^{r}(E)-\Sigma_{R}^{r}(E)\right]^{-1}
    \label{eq:5}
\end{equation}
and $G^{a}=[G^{r}(E)]^{\dagger}$, where $H_{C}$ is the Hamiltonian of the central UPM, the retarded self-energies $\Sigma_{L/R}^r(E)$ representing  the coupling with the leads can be calculated numerically by the recursive method \cite{ref68}.

In the following section, we investigate the conductance of the junction illustrated in Fig. \ref{fig:1} and its dependence on key parameters, including Fermi energy $E_F$, UPM strength $t_x$, junction length $L_x$, and interfacial barrier $\gamma$. By incorporating insights from the Fermi surface, dispersion relations, and conductance modulation mechanisms, we examine the conditions for realizing spin valves and spin transistors based on UPM junctions.

\begin{figure}[tbp]
    \centering
\includegraphics[width=\linewidth]{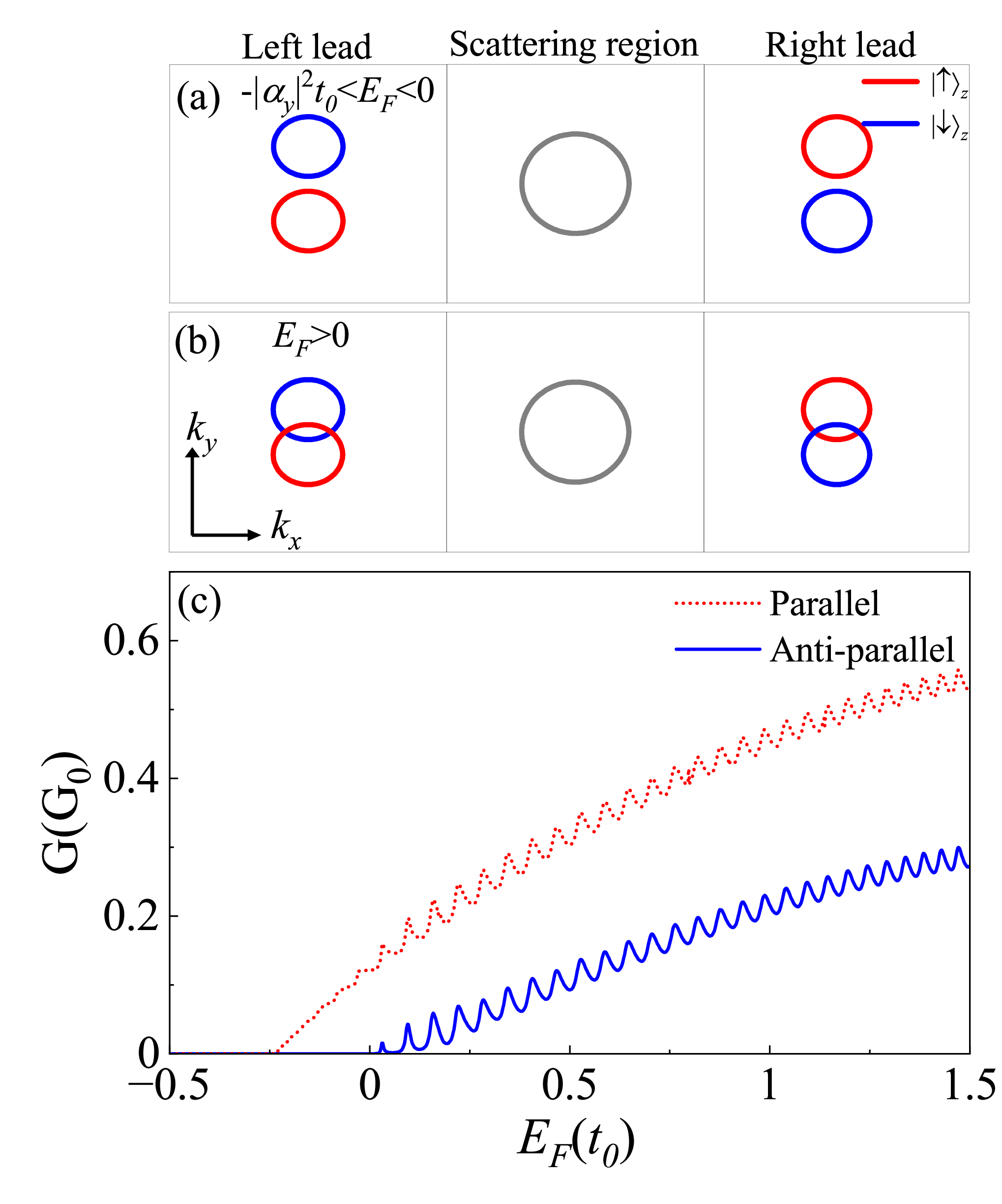}
    \caption{(a) and (b): Spin-split Fermi circles in two leads for the antiparallel configuration at negative and positive Fermi energies, respectively. (c): The conductance as the function of the Fermi energy of the UPM-based spin valve in the parallel and antiparallel configurations.
    The parameters used are as follows: $\alpha_{yL}=0.5$, $\alpha_{yR}=-\alpha_{yL}$, $\alpha_y=|\alpha_{yL}|=|\alpha_{yR}|$, \(U_0 = -2t_0\), \(L_x = 100a\), with $a=1$ the lattice constant and $t_0=1$ the energy unit.
    }
    \label{fig:2}
\end{figure}

\section{Results and Discussions}\label{Results and Discussions}

\begin{figure}[btp]
    \centering
\includegraphics[width=\linewidth]{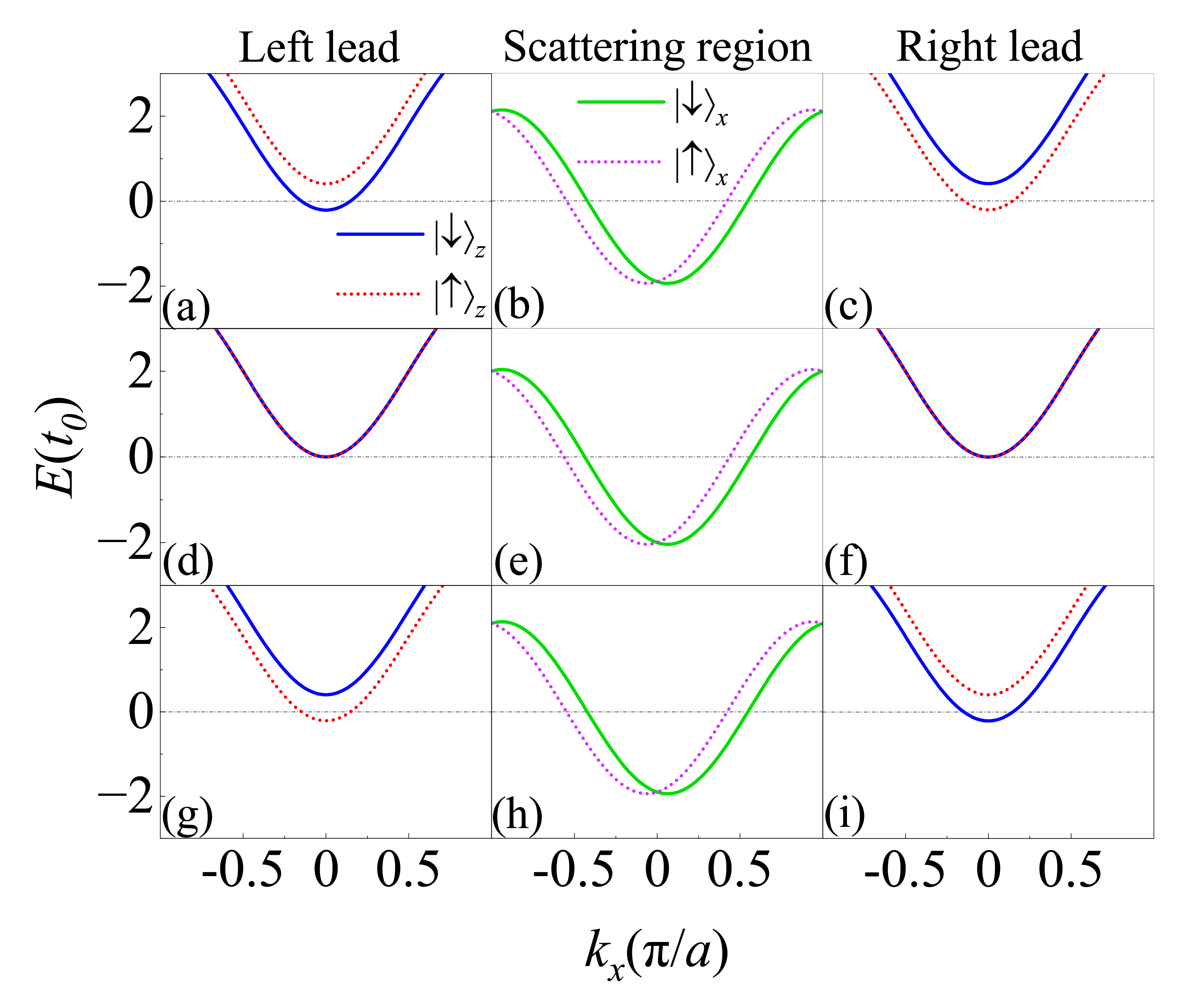}
    \caption{Spin-split subbands in the three regions of the UPM-based SFET in the antiparallel configuration of two leads. Panels (a)--(c), (d)--(f), and (g)--(i) correspond to $k_y = 0.1\pi/a$, $k_y = 0$, and $k_y = -0.1\pi/a$, respectively.
     Different colors correspond to distinct spin orientations. The parameters used are as follows: \(U_0 = -2t_0\), \(t_{yL} = 0.5t_0\), \(t_{yR} = -0.5t_0\), \(t_x = 0.2t_0\).}
    \label{fig:3}
\end{figure}

\begin{figure*}[tbp]
    \centering
\includegraphics[width=\textwidth]{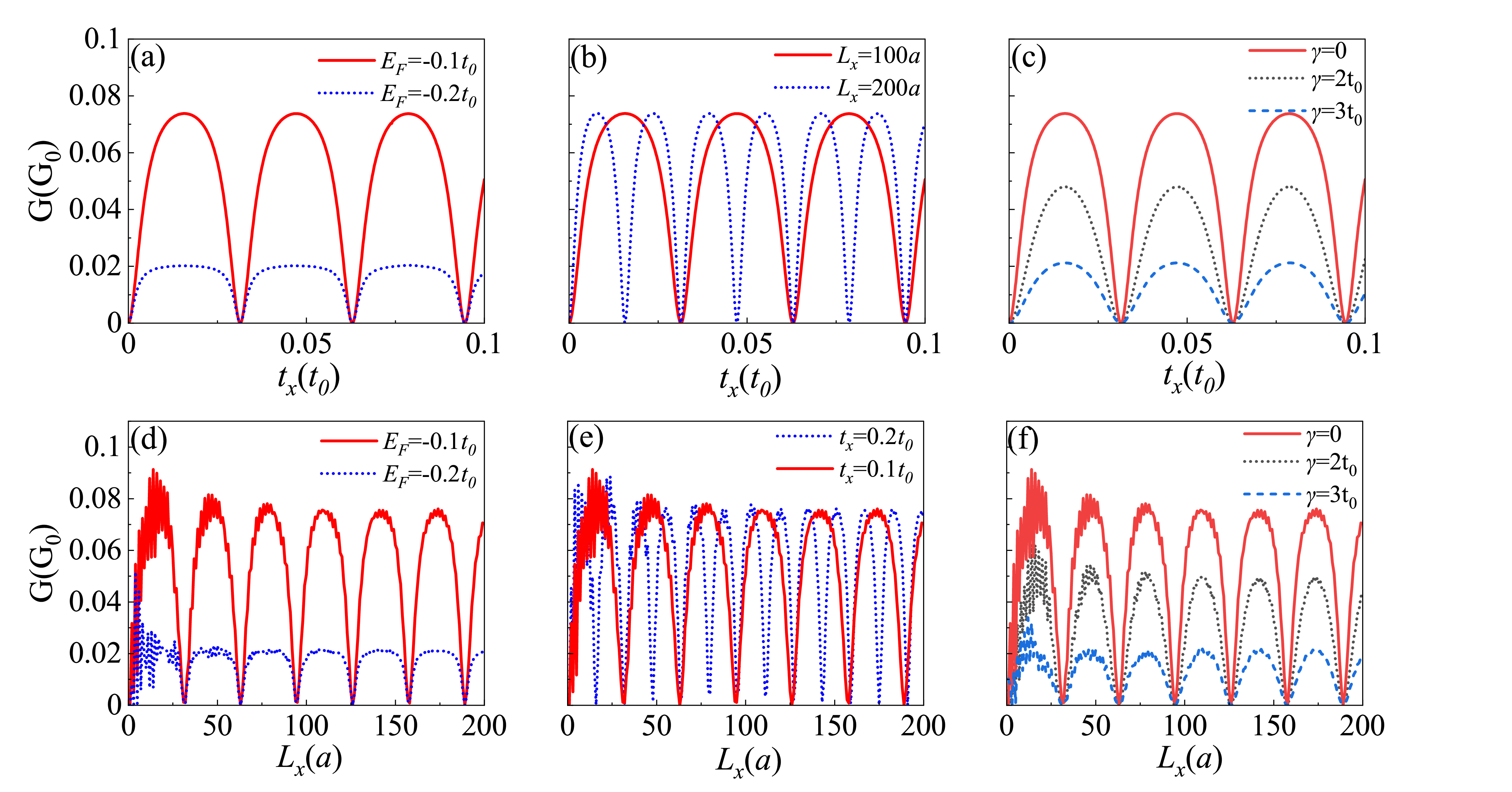}
    \caption{Conductance oscillations versus $t_x$ and $L_x$ at negative Fermi energies in the SFET. Parameters for (b) and (e): \(E_F = -0.1t_0\); for (c) and (f): \(E_F = -0.1t_0\), \(L_x = 100a\), \(t_x = 0.1t_0\); for all: \(U_0 = -2t_0\).}
    \label{fig:4}
\end{figure*}

In this section, we propose two electrically controlled spintronic devices that utilize the anisotropic spin splitting of UPMs: a spin valve (see Fig. \ref{fig:2}) and a spin transistor (see Figs. \ref{fig:3} and \ref{fig:4}). The dependence of the conductance on the strength vectors and the spin polarizations of UPMs enables tunable spin and charge transport in UPM-based junctions, offering a promising platform for zero-net-magnetization spintronic applications.

We start by considering the Hamiltonians of three regions of the device. By applying Fourier transforms, the low-energy Hamiltonian for the left and right leads in the momentum space can be obtained as
\begin{equation}
   H_{L/R}(\bm{k})=t_{0}(k_{x}^2+k_{y}^2)+2t_{yL/R}k_{y}\sigma_{z},
\label{eq:7}
\end{equation}
where $t_{yL/R}$ denotes the strength of $p_y$-wave magnetization in the left (right) lead and the lattice constant is set as $a=1$.
The Hamiltonian for the central UPM is
\begin{equation}
   H_{C}(\bm{k})=t_{0}(k_{x}^2+k_{y}^2)+2t_{x}k_{x}\sigma_{x}+U_{0}.
\end{equation}
Note that the exchange field in UPMs can be consistently defined as $(\bm{t}\cdot\bm{k})\bm{\sigma}\cdot\bm{n}$, where $\bm{t}$ denotes the strength vector and $\bm{n}$ represents the spin polarization vector.

The spin-resolved band structures for the two UPM leads can be found as
\begin{equation}
   E_{L/R}=t_0\left[k_{x}^{2}+(k_{y}+\sigma\alpha_{yL/R})^{2}-\alpha_{yL/R}^{2}\right],
\label{eq:10}
\end{equation}
where $\alpha_{yL/R}=t_{yL/R}/t_{0}$, and $\sigma=1 (-1)$ for the spin-up (-down) subband with respect to the $z$ direction. For the central UPM, the band dispersion is
\begin{equation}
   E_{C}=t_0\left[(k_{x}+\sigma'\alpha_{x})^{2}+k_{y}^{2}-\alpha_{x}^{2}\right]+U_{0},
\label{eq:11}
\end{equation}
where $\alpha_{x}=t_{x}/t_{0}$, and $\sigma'=1 (-1)$ for the spin-up (-down) subband with respect to the $x$ direction.

First, we consider a UPM/normal-metal/UPM junction which acts as a spin valve. To model the central region as a normal metal, we set $t_x=0$. According to Eq. (\ref{eq:10}), the left and right leads exhibit spin-split Fermi circles at a fixed Fermi energy $E_F$, with the splitting occurring along the $k_y$ direction, as illustrated in Fig. \ref{fig:2}(a) and (b). When $E_F>0$, the two spin-split Fermi circles still overlap. In contrast, for $E_F<0$, two circles become completely separated. It is important to note that both the transverse momentum $k_y$ and the spin component $\sigma_z$ are conserved during transport when $t_x=0$. Consequently, a transport channel opens only when both $k_y$ and the spin state are matched between the two leads.

We examine both parallel ($t_{yR}=t_{yL}$) and antiparallel ($t_{yR}=-t_{yL}$) configurations of the strength vectors in the two UPM leads. In the antiparallel configuration, when the Fermi energy lies within the interval $E_F \in (-\alpha_{y}^{2}t_{0}, 0)$, with $\alpha_y=|\alpha_{yL}|=|\alpha_{yR}|$, the two Fermi circles are spin-split in opposite directions along the $k_y$ axis and are fully separated, as shown in Fig. \ref{fig:2}(a). For a fixed $k_y$, the spin orientations in the two leads are opposite. This spin mismatch prevents the opening of any transport channel. Incident electrons from the left lead  are thus completely reflected, resulting in zero conductance, as depicted in Fig. \ref{fig:2}(c). However, when $E_F > 0$, a partial overlap occurs between two spin-split Fermi circles in both leads, as shown in Fig. \ref{fig:2}(b). Within this overlapping $k_y$ region, both spin-up and spin-down states are available in both leads, allowing transport channels to open and yielding a non-zero conductance. As the Fermi energy increases, the overlap area of the Fermi circles expands, leading to a general increase in conductance, albeit with slight oscillations caused by multiple reflections at the interfaces \cite{ref43,Lv2025Gate}.

Conversely, in the parallel configuration, the Fermi circles in both leads are split in the same direction along the $k_y$ axis. For any given $k_y$, the spin states are identical in both leads. Transport channels are open whenever the Fermi energy is above the bottom of the subbands, i.e., $E_F>-\alpha_{y}^{2}t_{0}$.
The significant contrast in conductance between the antiparallel and parallel configurations within the energy range $E_F \in (-\alpha_{y}^{2}t_{0}, 0)$ enables a high on/off ratio in this UPM-based spin valve, which can be electrically controlled by tuning the strength vectors of the leads.

Unlike conventional ferromagnetic spin valves \cite{DalDin2024Antiferromagnetic,ref70}, the proposed UPM spin valve exhibits zero net magnetization. Its conductance is governed by the relative alignment of the UPMs' strength vectors rather than macroscopic magnetizations. More importantly, switching between parallel and antiparallel configurations in conventional spin valves typically requires an external magnetic field to reverse the magnetization of one ferromagnetic lead. In contrast, switching in our UPM-based spin valve is achieved by electrically tuning the strength vectors, offering a more practical control mechanism.

Second, we consider a UPM/UPM/UPM junction functioning as a SFET. Here, the central normal metal is replaced by a UPM whose strength vector and spin quantum axis are both along the $x$ direction. The perpendicular orientation of this spin axis relative to those in the leads, combined with the longitudinally split Fermi circles (see Fig. \ref{fig:1}), enables spin precession for incident electrons traversing the central region. Based on the tight-binding Hamiltonians, we plot the spin-split band dispersions for each region of the SFET in Fig. \ref{fig:3}, illustrating the preserved time-reversal symmetry. In the antiparallel configuration with $E_F<0$, the spins of the occupied subbands are opposite in the two leads, which would result in zero conductance in a simple spin valve junction. However, the inclusion of the central $p_x$-wave UPM, with its longitudinally split subbands and perpendicular spin axis (see Fig. \ref{fig:3}), induces spin precession. According to Eq. (\ref{eq:11}), the wave vector difference is $\Delta k_x=2\alpha_x$, leading to a precession angle of $2\alpha_xL_x$. The conductance reaches a maximum \cite{Ciorga_2025} when $2\alpha_xL_x=(2n+1)\pi$ and drops to zero when $2\alpha_xL_x=2n\pi$, where $n$ is an integer.

Fig. \ref{fig:4} demonstrates periodic oscillations in conductance as a function of the UPM strength $t_x$ and the length $L_x$ of the central region. For $E_F<0$, the conductance minima can reach zero due to perfect spin precession within the central UPM. A key advantage is that the wave vector difference $\Delta k_x=2\alpha_x$ remains constant for all transverse modes $k_y$, a consequence of the longitudinal spin-splitting along the $k_x$ direction. This uniformity ensures that all transverse modes precess at the same frequency, allowing their transmissions to vanish simultaneously and resulting in a true zero conductance state -- a crucial feature for a high-performance SFET with a high on/off ratio.
This behavior contrasts with the original Datta-Das transistor based on Rashba spin-orbit coupling \cite{Datta1990Electronic}, where differing precession frequencies for different transverse modes prevent the conductance from reaching zero in the off state. As shown in Fig. \ref{fig:4}(b), doubling the length $L_x$ halves the oscillation period of the conductance versus $t_x$, which is consistent with the spin precession analysis. Notably, $t_x$ can be tuned effectively by modulating the spin polarization via an external electric field \cite{Song2025Electrical}, providing electrical control of the SFET.

In addition to the primary spin precession oscillations, the conductance versus $L_x$ also exhibits minor oscillations due to multiple reflections at the interfaces, as shown in Figs. \ref{fig:4}(d)-(f). These resonance transmission induced oscillations are absent in the dependence versus $t_x$ dependence because varying $t_x$ simultaneously alters both the right-going and left-going wave vectors by the same amount, leaving the resonant conditions relatively unaffected.

Finally, we examine the effect of interfacial barriers on the performance of SFET. Figs. \ref{fig:4}(c) and (f) show the conductance for varying barrier strength $\gamma$. While stronger interfacial barriers reduce the maximum conductance, they also smoothen the conductance minima, which can enhance the practical tunability of the device. The introduction of inelastic impurities in the central scattering region produces an effect similar to that of interfacial barriers on the conductance, leading to a moderate suppression of the conductance maxima and a concomitant smoothing of the conductance minima, with results analogous to those presented in Figs. \ref{fig:4}(c) and (f).

 \begin{figure}[tbp]
	\centering
	\includegraphics[width=\linewidth]{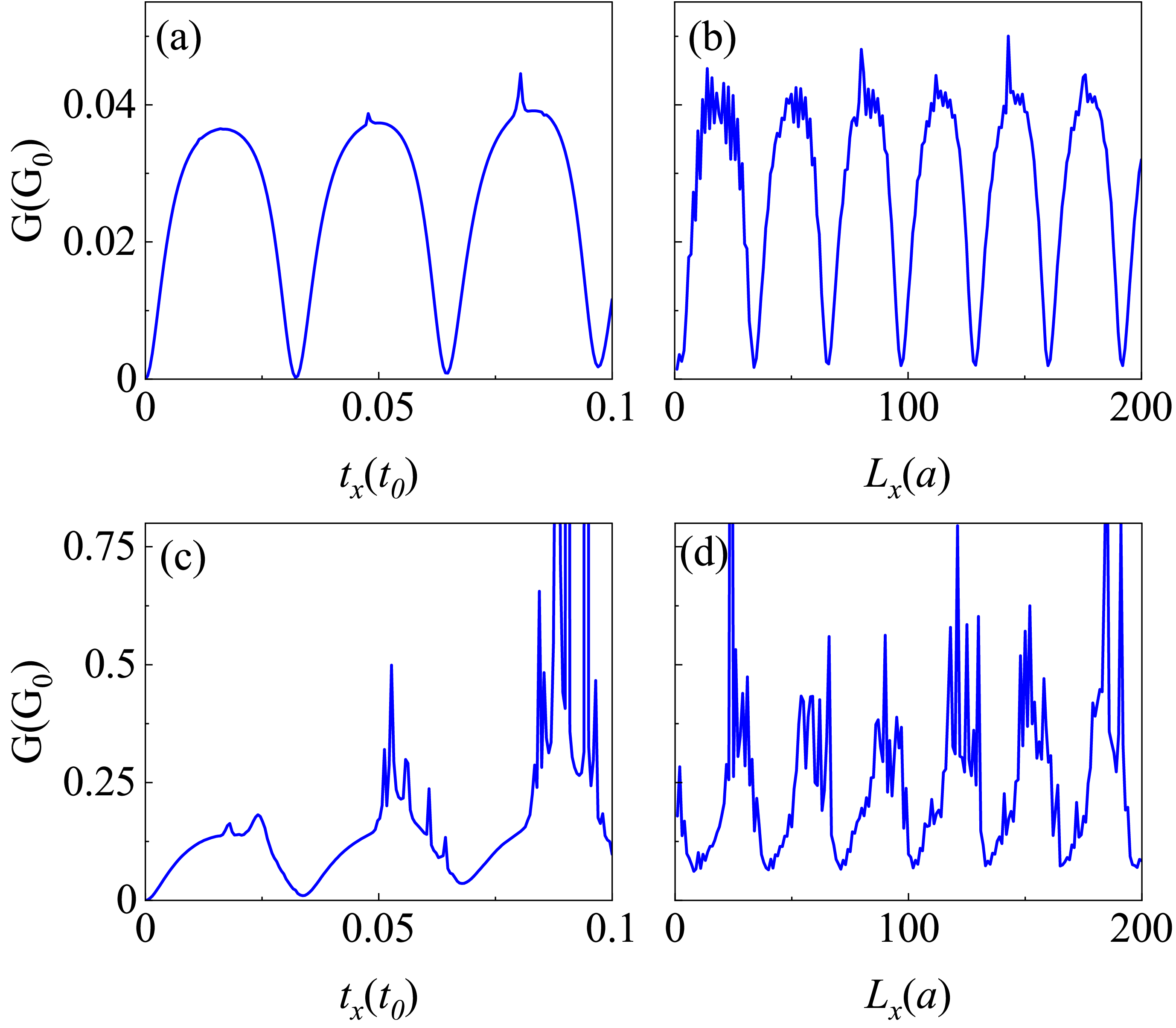}
	\caption{Influence of the buffer layer on the conductance of the SFET. (a) and (b) show the periodic conductance for a buffer layer width of $L_n = 3a$, while (c) and (d) correspond to $L_n = 10a$, where $a$ is the lattice constant. The other parameters are set as follows: for (a) and (c), $L_x = 100a$; for (b) and (d), $t_x = 0.1t_0$; and $E_F = -0.1t_0$ for all subfigures.}
	\label{fig:6}
\end{figure}

To reduce the difficulty of controlling spin polarization in a specific layer of the heterostructure and to mitigate the influence of the applied local electric field on neighboring layers, we introduce buffer layers of width $L_n$ between the central scattering region and the two leads. As shown in fig. \ref{fig:6}, the conductance exhibits a distinct dependence on the buffer layer width. For a narrow buffer layer (e.g., $L_n = 3a$), the maximum conductance is reduced compared to the case without a buffer layer, yet a high on/off ratio is preserved. This indicates that even a buffer layer of limited width can effectively suppress mutual interference between neighboring layers. In contrast, for a wider buffer layer (e.g., $L_n = 10a$), the conductance increases substantially, enhancing the on-state performance, albeit at the cost of a slightly reduced on/off ratio. These results demonstrate that the buffer layer design provides a tunable trade-off between minimizing interlayer crosstalk and optimizing conductance, thereby effectively reducing the difficulty of selectively tuning spin polarization in a given layer while also limiting the impact of local electric fields on adjacent regions.

\section{Conclusion}\label{Conclusion}

This work proposes two core spintronic devices, a spin valve and a spin transistor,  based on UPMs. Both devices operate under time-reversal symmetry and require neither net magnetization nor relativistic spin-orbit coupling, offering a distinct mechanism for spin manipulation.
The spin valve effect is achieved by electrically switching the strength vectors of two UPM electrodes between parallel and antiparallel configurations. Conductance is suppressed in the antiparallel state when two Fermi circles separate in momentum space, yielding a high on/off ratio. Replacing the central normal region with an orthogonally polarized UPM converts the device into a spin transistor. The longitudinal spin-splitting in the central region enables uniform spin precession for all transverse modes, leading to full-conductance oscillations and a perfectly modulated on/off state -- a key advantage over conventional spin-orbit-coupled transistors. Electrical tunability of the UPM strength vectors allows active control of conductance, as demonstrated by oscillation periods dependent on $t_x$ and $L_x$. Supported by recent experiments on electrical switching in UPMs, these findings establish UPMs as an integrable platform for low-power, magnetic-field-free spintronic applications \cite{Song2025Electrical}.

\section*{Data Availability}
The data supporting the findings of this study are available from the corresponding authors upon reasonable request.

\section*{Author Contributions}
Z.-Y. Y. developed the theoretical model, performed calculations, and wrote the manuscript.
J.-F. L. and J. W supervised the project.
P.-H. F provided valuable insights and contributed to the analysis and interpretation of the results.
All authors discussed the results and contributed to the final version of the manuscript.

\section*{Competing Interests}
The authors declare no competing interests.

\begin{acknowledgments}
The work described in this paper is supported by the National Natural Science Foundation of China (Grants No. 12174077 and No. 12174051).
\end{acknowledgments}

\end{document}